# Theoretical study of subwavelength imaging by acoustic metamaterial slabs

Ke Deng[1,2], Yiqun Ding[1], Zhaojian He[1], Heping Zhao[2], Jing Shi[1], and Zhengyou Liu[1,a)]

[1]Key Lab of Acoustic and Photonic materials and devices of Ministry of Education and Department of Physics, Wuhan University, Wuhan 430072, China

[2]Department of Physics, Jishou University, Jishou 416000, Hunan, China

We investigate theoretically subwavelength imaging by acoustic metamaterial slabs immersed in the liquid matrix. A near-field subwavelength image formed by evanescent waves is achieved by a designed metamaterial slab with negative mass density and positive modulus. A subwavelength real image is achieved by a designed metamaterial slab with simultaneously negative mass density and modulus. These results are expected to shed some lights on designing novel devices of acoustic metamaterials.

[a)]To whom all correspondence should be addressed, e-mail address is zyliu@whu.edu.cn

# I. INTRODUCTION

Recent advances in electromagnetic (EM) metamaterials (MMs)[1] provide the foundation for realizing many intriguing phenomena, such as inverse Doppler Effect,[2] negative refraction,[3] and amplification of evanescent waves.[4] These phenomena can be utilized to design novel EM devices. Pendry found that, as the combined result of negative refraction and amplification of evanescent waves, a MM slab with effective permittivity $\varepsilon = -1$ and permeability $\mu_{EM} = -1$ can focus both the propagating and evanescent waves of a point source into a perfect image.[5] Thereby such a slab device has been referred as the perfect lens. Pendry's perfect lens stimulated lots of research interests due to the great significance of subwavelength imaging in various applications (see the ref. 6 and references therein).

Acoustic MMs have also attracted much attention more recently.[7-14] Till now most concerns have been focused on their realization. MMs with negative effective mass density has been realized by a class of three-component phononic crystal,[7,8] and more recently by a membrane-type structure.[9] An ultrasonic MM consisting of arrays subwavelength Helmholtz resonators has been demonstrated to exhibit a negative effective bulk modulus.[10] MMs with simultaneously negative mass density and bulk modulus have also been realized with liquid[11] and solid[12] matrix. These offer promising opportunities to design novel acoustic devices.[13,14] However, studies on such subjects are still relatively lacking.

Recently, Zhang *et al* investigated the surface waves between a fluid half-space and an acoustic MM half-space.[13] They concluded that the negative mass density is the necessary condition for the existence of surface states on acoustic MMs. In addition, a super lens formed by a hypothetical liquid-based MM slab with negative mass density and positive bulk modulus was proposed. In this paper, we investigate the subwavelength imaging by solid-based acoustic MM slabs immersed in the liquid matrix. Two kinds of imaging are considered. In the first one, we design a MM slab with negative mass density and positive modulus. In this single-negative (SN) case, a near-field subwavelength image formed by evanescent waves is obtained as the result of the coupling of surface waves at two interfaces of the slab. In the second one, we design a MM slab with simultaneously negative mass density and modulus. In this double-negative (DN) case, a subwavelength real image is obtained as the combined effects of negative refraction and amplification of evanescent waves. In addition, it is demonstrated that our results on the subwavelength imaging will still remain valid when the material absorptions are considered.

Analysis in this paper is based on the effective-medium description[15] of MMs. In this description, MMs can be treated as dispersive homogeneous materials with negative effective parameters, because their structure constants are much small compared with the free-space wavelength at the operating frequency. We will consider a solid MM slab with effective mass density $\rho_{eff}(\omega)$, effective modulus $E_{eff}(\omega)$ and $\mu_{eff}(\omega)$, immersed in an ordinary liquid matrix with mass density $\rho_1$ and modulus $E_1$. Here, $E \equiv \lambda + 2\mu$ is the elastic constant which governs the velocity of the longitudinal wave by $C_l = \sqrt{E/\rho}$, $\lambda$ and $\mu$ are Lamè's constants, and $\mu$ governs the velocity of the transverse wave by $C_t = \sqrt{\mu/\rho}$. The effective parameters of our designed MMs will be calculated by the CPA method.[15] Other results in this paper will be calculated by a standard transfer matrix technique.

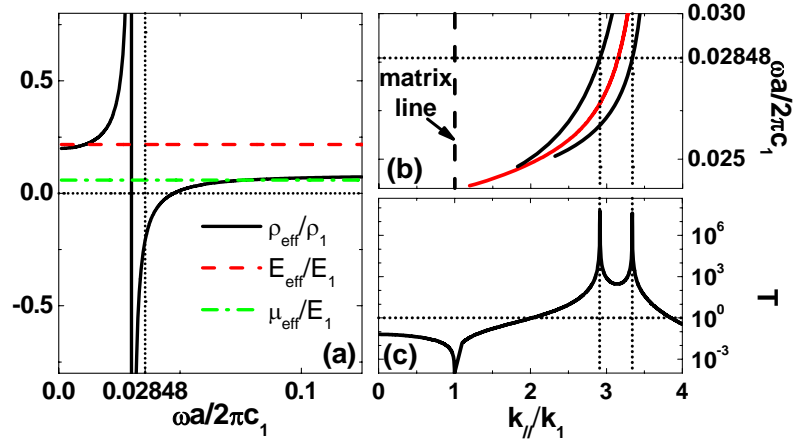

FIG. 1. (Color online) (a) Effective parameters of the designed SN MM derived from the CPA method; (b) Surface wave dispersion curves for the system of Fig.1(a). The middle red (grey) line represents the surface waves between a liquid half-space and a MM half-space. The other two black (dark) lines denote the symmetric and antisymmetric surface modes of a MM slab as the results of the coupling of surface waves at two single interfaces. Here the lab thickness is $d = 0.2a/0.02848$; (c) Transfer function of the MM slab at the operating frequency $0.02848$

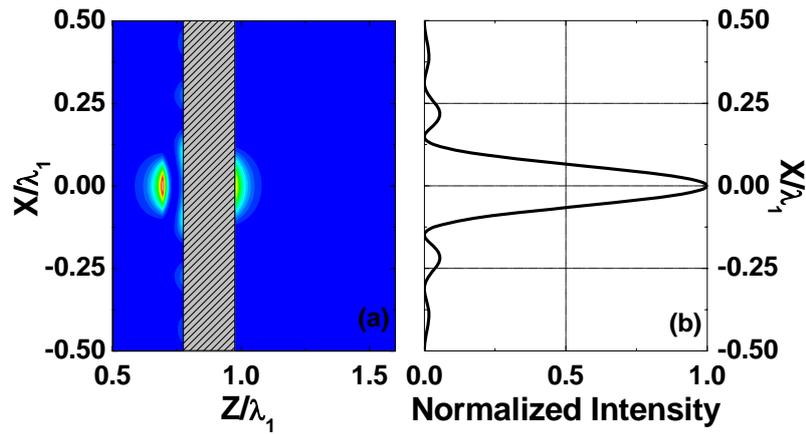

FIG. 2. (Color online) Imaging results for the SN slab: (a) Pressure field intensity distribution in the liquid; (b) Transverse field distribution at the image plane.

## II. SINGLE NEGATIVE IMAGING

To realize the SN MMs, we design a structure consisting of an fcc array of rubber-coated gold spheres in epoxy, with a filling fraction of 9.77%. The rubber coating has an inner-to-outer radius ratio of 17/18. This kind of three-component phononic crystal has been demonstrated to exhibit a negative effective mass density in our previous works.[7,8,12] We use the CPA method[15] to estimate the effective parameters for the present system. Figure 1(a) shows the $\rho_{eff}$, $E_{eff}$ and $\mu_{eff}$ versus the normalized frequency $\omega a/2\pi c_1$, where $a$ is the lattice constant and $c_1$ is the wave velocity in the liquid. Here we have chosen the liquid matrix as mercury. The material parameters used in the above calculations are for gold: $\rho = 19500.0\, kg/m^3$, $E = 2.2 \times 10^{11}\, N/m^2$, $\mu = 2.99 \times 10^{10}\, N/m^2$; for rubber: $\rho = 1300.0\, kg/m^3$, $E = 1.475 \times 10^6\, N/m^2$, $\mu = 3.25 \times 10^4\, N/m^2$; for epoxy: $\rho = 1180.0\, kg/m^3$, $E = 7.61 \times 10^9\, N/m^2$, $\mu = 1.59 \times 10^9\, N/m^2$; and for mercury: $\rho = 13500.0\, kg/m^3$, $E = 2.84 \times 10^{11}\, N/m^2$. Fig. 1 (a) shows that a SN systems is obtained in the frequency range $[0.02398, 0.03785]$. In this SN frequency range, a common gap opens for both the longitudinal and the transverse waves, thus most of the propagating waves emitted by the source placed in front of the slab are reflected at the first interface. At the same time, surface waves coupled at the two interfaces, which are stimulated by evanescent waves of the source, bring its subwavelength details to the back side of the slab. Therefore a near-field subwavelength image formed by evanescent waves can be obtained. The key in SN imaging is to delicately design the system so that the slab possesses a SN frequency range and, in the same range, a surface wave dispersion curve which deviates from the matrix-line as far as possible. Three surface wave dispersion curves for the system of Fig. 1 (a) are plotted in Fig. 1(b). The middle red (grey) line is for the surface waves between a liquid half-space and a MM half-space. The other two black (dark) lines denote the symmetric and antisymmetric surface modes of a MM slab as the results of the coupling of surface waves at two single interfaces. The transfer function $T$, which is defined as the ratio of amplitude of pressure fields across the slab, for the MM slab of thickness $d = 0.2\lambda_1$ at operating frequency $\omega a/2\pi c_1 = 0.02848$ is plotted in Fig.1(c). Here $\lambda_1$ is the wave length in the liquid matrix. One sees from Figure 1(c) that evanescent waves are enhanced in the range of $k_p \in [2.0, 3.9]$. Here $k_p \equiv k_{//}/k_1$ is the normalized transverse wave number, where $k_{//}$ denotes the transverse wave number and $k_1$ is the wave number in the liquid. The two peaks at $k_p = 2.9$ and $k_p = 3.3$ respectively are exactly consistent with the corresponding modes at the operating frequency in Fig 1(b). These enhancements give the possibility of restoring the evanescent components of the source, thus leading to the realization of superlensing. In Fig 2(a), we give the pressure field intensity of the system with a point source placed in front of the slab with a distance $d_s = 0.08\lambda_1$. Fig. 2(b) shows the transverse field distribution at the image plane. It's very clear to see from Fig. 2 that a near-field subwavelength image with a resolution of $0.13\lambda_1$ is achieved. It should be pointed out here that since this image is formed completely by evanescent waves, there is no fixed focal plane. Here, the right interface of the slab in the liquid is considered as the image plane.[13] Due to the single negative nature of the slab, the intensity of image is much less than the source as one can see from Fig. 2(a).

## III. DOUBLE NEGATIVE IMAGING

For the realization of DN MMs, we combine an array of bubble-contained water spheres with a relatively shifted array of rubber-coated gold spheres in epoxy matrix to form a zinc blende structure. The filling fraction of the first fcc lattice is 26.2% and the ratio of the radii of the air bubble to the water sphere is $2/25$. The filling fraction of the second fcc lattice is 9.77% and the rubber coating has an inner-to-outer radius ratio of 15/18. Such a structure has been proved to exhibit simultaneously negative effective mass density and bulk modulus in our recent paper.[12] Figure 3(a) shows the effective parameters of this system calculated by the CPA method. Here we have chosen the liquid matrix as mercury. The material parameters used in the above calculations are for air: $\rho = 1.23\,kg/m^3$, $E = 1.42\times10^5\,N/m^2$; for rubber: $\rho = 1300.0\,kg/m^3$, $E = 2.213\times10^9\,N/m^2$, $\mu = 9.98\times10^6\,N/m^2$; and for water: $\rho = 1000\,kg/m^3$, $E = 2.22\times10^9\,N/m^2$. As one sees from Fig 3(a), a DN system is obtained in the frequency range $[0.28308, 0.30421]$. Surface wave dispersion curves for this system are plotted in Fig. 3(b). Physical meanings of the three lines in Fig. 3(b) are the same as those of Fig. 1(b). The operating frequency is set as $\omega a/2\pi c_1 = 0.29558$ at which $\rho_{eff} = -0.75334\rho_1$ and $E_{eff} = -0.8108 E_1$. Transfer function of the slab with a thickness $d = 0.5\lambda_1$ at the operating frequency is plotted in Fig. 3(c). The imaging results with a point source placed in front of the slab with distance $d_s = 0.25\lambda_1$ are given in Fig. 4. Here, Fig. 4(a) gives the phase distribution and Fig. 4(b) gives the intensity distribution. As one can see that a real image is obtained behind the slab which is lacked in the SN case. This image is formed as a result of negative refraction. With simultaneously negative $\rho_{eff}$ and $E_{eff}$, propagating parts of the source are negatively refracted in the form of longitudinal waves in the slab and refocused at the image plane behind the slab. The inset of Fig 4(a) gives the phase distribution in the slab. As one can see, similar to the EM case,[5] there is a first focus in the slab. At the same time, coupled surface waves at two interfaces bring subwavelength details from the source to the image plane. Therefore a subwavelength real image is obtained. We calculated the field distribution at the image plane and gained a $0.35\lambda_1$ resolution as shown in Fig 4(c).

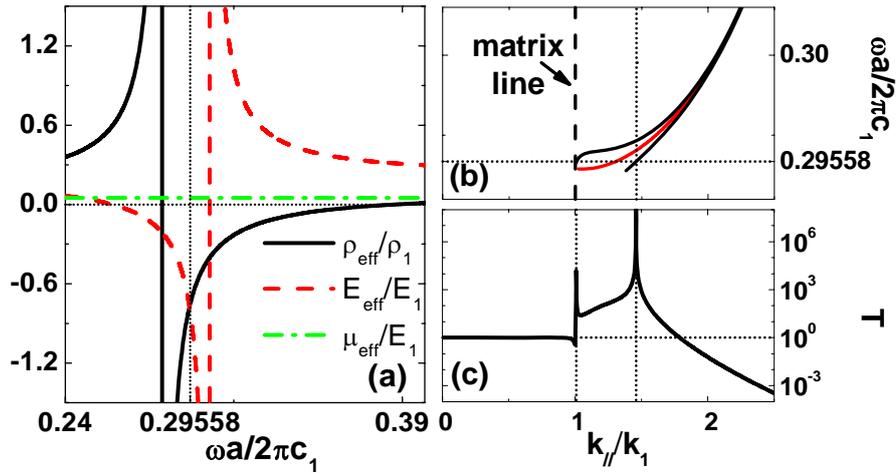

FIG. 3. (Color online) (a) Effective parameters of the designed DN MM derived from the CPA method; (b) Surface wave dispersion curves for the system of Fig.3(a). The middle red (grey) line represents the dispersion curve for the surface waves between a liquid half-space and a MM half-space. The other two black (dark) lines denote the symmetric and antisymmetric surface modes of the MM slab as the results of the coupling of surface waves at two single interfaces. Here the slab thickness is $d = 0.5a/0.029558$; (c) Transfer function of the MM slab at the operating frequency $0.029558$.

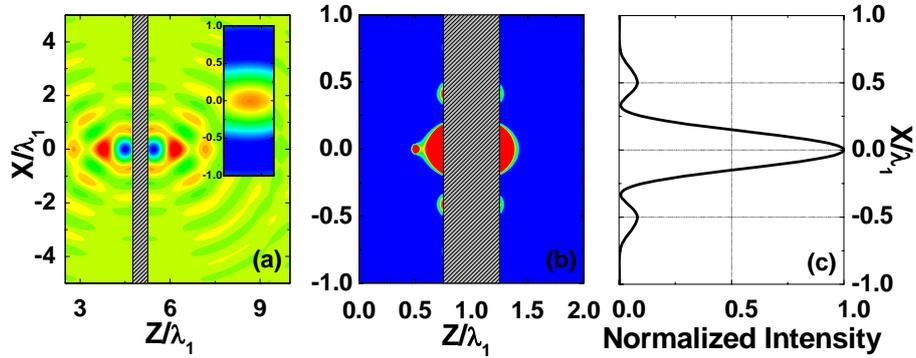

FIG. 4. (Color online) Imaging results for the DN slab: (a) Phase distribution in the liquid. Inset gives the phase distribution in the slab; (b) Intensity distribution in the liquid; (c) Transverse field distribution at the image plane.

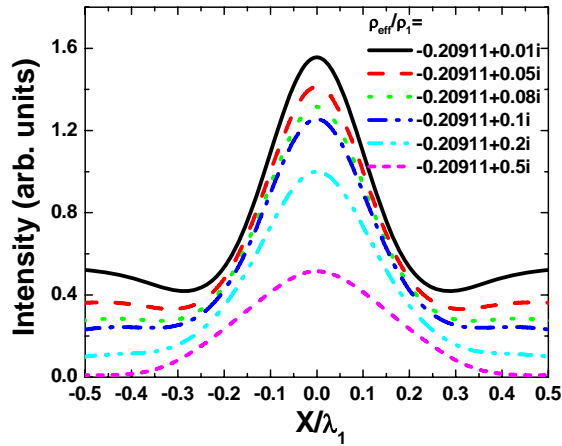

FIG. 5. (Color online) The transverse intensity distribution for imaging with material absorption. The slab and point source are otherwise identical to those in Fig. 2.

## IV. EFFECTS OF ABSORPTION

Till now we have investigated the SN and DN imaging in an ideal situation with no material absorption. In reality, material losses are always present and will generally weaken the effect of superlensing. However, it can be expected that, just as in the EM case,[6] in the limit of small material loss, our results on the subwavelength imaging will still remain valid. As an example, we consider the same system discussed in Fig. 2, but alter the effective mass density from $\rho_{eff} = -0.20911\rho_1$ to $\rho_{eff} = (-0.20911 + bi)\rho_1$. The imaginary part $b$ is varied from $0.01$ till $0.5$. As shown in Fig. 5, with the increase of the absorption, the intensity of the transmitted fields is attenuated, and the subwavelength characters in the transverse field distribution gradually disappear. Nevertheless, the subwavelength imaging is still achieved even if the imaginary part of $\rho_{eff}$ reach the same order of its real one.

## V. SUMMARY

In summary, two kinds of subwavelength imaging by acoustic MM slabs immersed in the liquid matrix have been investigated. In the first one, a MM slab with negative mass density and positive modulus was designed to achieve a near-field subwavelength image formed by evanescent waves. In the second one, a MM slab with simultaneously negative mass density and modulus was designed to achieve a subwavelength real image. Our results are expected to shed some lights on designing novel application devices of acoustic MMs.

## ACKNOWLEDGMENTS

This work is supported by the National Natural Science Foundation of China (Grant Nos. 10874131, 10731160613, 50425206 and 10418014).